\documentclass{aa}  
\usepackage{graphicx}
\usepackage{txfonts}
\usepackage{longtable}
\usepackage{hhline}
\usepackage{pdflscape}
\newcommand{\te }{\mbox{$T_{\rm eff}$}}

\newcommand{\be}{\begin{equation}}
\newcommand{\ee}{\end{equation}}

\newcommand{\feh}{\hbox{$ [{\rm Fe}/{\rm H}]$ }}
\def\c2{\chi ^2}

\begin{document}
\title
{Identification of metal-poor stars using the artificial neural network
\thanks{Table~1, Table~2, Figure 2, Figure 3, and Figure 6 are available in electronic form  via
http://www.edpsciences.org}
}
\author
{ Sunetra Giridhar, \inst{1}
 Aruna Goswami, \inst{1}
Andrea Kunder, \inst{2}
 S. Muneer, \inst{3} \and
G. Selvakumar, \inst{4}
}
\offprints { Sunetra Giridhar}

\institute{Indian Institute of Astrophysics, Koramangala, Bangalore 560034, India\\
\email{giridhar@iiap.res.in, aruna@iiap.res.in}
\and
Cerro Tololo Inter-American Observatory, NOAO, Casilla 603, La Serena, 
Chile\\
\email{akunder@ctio.noao.edu}
\and
CREST Campus, Indian Institute of Astrophysics, Hosakote 562114, India\\ 
\email{muneers@iiap.res.in}
\and
Vainu Bappu Observatory, Indian Institute of Astrophysics, Kavalur, 
635701, India\\ 
\email{selva@iiap.res.in}}

   \date{}

 
  \abstract
   { Identification of metal-poor stars among field stars  is extremely useful
for studying the structure and evolution of the Galaxy and of external
galaxies.}
    {We search for metal-poor stars using the artificial neural network (ANN) and extend its usage 
    to determine absolute magnitudes.}  
   {We have constructed a library of 167 medium-resolution stellar spectra (R $\sim$ 1200) 
    covering the stellar temperature range of 4200 to 8000 K, log~$g$ range of 0.5 to 5.0, and
    [Fe/H] range of $-$3.0 to $+$0.3 dex. This empirical spectral library was 
     used to train ANNs, yielding  an accuracy of 0.3 dex in \feh, 200 K in
 temperature, and 0.3 dex in log~$g$.  We found that the independent  
 calibrations of near-solar metallicity stars and metal-poor stars
 decreases the errors in  \te~ and log~$g$~ by nearly a factor of two.}
   {We calculated  \te, log~$g$, and  \feh on a consistent
 scale for a large number of field stars and candidate metal--poor
 stars.  We extended the application of this method to
 the calibration of absolute magnitudes using nearby stars with well-estimated 
parallaxes. A better calibration accuracy for
 M$_V$ could be obtained by training separate ANNs for cool,
  warm, and metal-poor stars. The current accuracy of M$_{V}$ calibration
  is $\pm$0.3 mag.}
   { A list of newly identified metal-poor stars is presented.
     The M$_{V}$ calibration procedure developed  
    here is reddening-independent and hence may serve as a powerful
     tool in studying galactic structure.}

   \keywords{stars: individual -- stars: solar-type -- stars: metal-poor--- stars: fundamental parameters
               }
\titlerunning{Application of Artificial Neural Network for Stellar parametrization}
\authorrunning{Giridhar et al.}
   \maketitle
%

\section{Introduction}

  Metallicity estimates  for large samples of stars among 
 different Galactic components can provide a wealth of information
 on the structure and formation of our Galaxy.  Extremely metal-poor
 stars are the relics of early the Galaxy,  while moderately metal-poor
 stars can provide indications of whether it is a thick or thin disk when supplemented by
  additional information  such as the kinematics of these objects.
 A high spectral resolution follow-up of these
  metal-poor stars (identified mostly through
 low- and intermediate-resolution spectral surveys)
 has resulted in identifications of exotic objects such as
 very metal-poor (VMP), extremely metal-poor (EMP), ultra metal-poor 
(UMP), and hyper metal-poor (HMP) (explained in \cite{be05}), which
 show different degrees of metal deficiencies.
 Among these metal-poor class, subclasses comprising  
 carbon-enhanced metal-poor stars
 (CEMPs) have also been identified, which show a wide range in s- and r-process element 
enhancements. These objects are important tools
 for  understanding the  enrichment of the interstellar medium (ISM) caused by
 stars of different mass range in our  Galaxy.

 Intrinsic luminosity is another important parameter that 
 not only helps in deriving the distances of the objects,  but 
 also helps in distinguishing objects at different evolutionary stages.
 The photometric determination of M$_{V}$, however, requires a good reddening estimate.
 The spectroscopic approaches  based on line strengths, line
 ratio, and profiles of H {\sc i}, Ca {\sc ii}, etc. are reddening independent.
  A list of luminosity-sensitive
 features for different spectral types can be found in \cite{gr09},
  and a condensed review in \cite{gi10}.  

 A large number of metal-poor stars have been identified with
 the help of earlier surveys such as the HK Survey (\cite{be92}). However, 
 multi-object spectrometers like 6df 
   on the UK Schmidt telescope (\cite{wa98}), 
 AAOMEGA at the Anglo-Australian Telescope (AAT) (\cite{sh06}), and 
the LAMOST project (\cite{zh06}) can provide a large number of spectra 
per night. The ongoing and future surveys and space missions will collect 
a vast amount of spectra for  stars belonging to different components of 
our Galaxy and nearby galaxies.
 The wide variety of objects covered in these surveys require good pipelines
 for data handling and automated procedures that are efficient 
 as well as robust in deriving accurate stellar parameters that are
essential ingredients in studying the structure and evolution of our Galaxy.

  Several automated methods of spectral classification and 
parametrization such as the minimum distance method (MDM),
 the gaussian probabilistic model  (GPM), 
 the principal component analysis (PCA), and the neural network have been developed 
 over the last two decades. These 
 methods have been summarized in \cite{ba02}.

 These automated methods  differ in two major ways.
 Real stellar spectra of well-known calibrated stars,
 referred to as the empirical library, are employed by some groups (including us),
 while others prefer using a synthetic spectral library. Both approaches 
 have their merits and disadvantages. Synthetic spectra depend on the quality with which
 the model atmosphere (often assuming local thermodynamic equilibrium) represents actual stars, and
 the line lists used are sometimes poor; in particular the line data for
 molecular lines are not very accurate. Our attempt at validating these line lists
 by comparing the synthetic spectra with spectra of well-known stars has shown
 disagreements that indicate that there are unidentified lines or that the oscillator strengths
 of poor quality. The problem is more severe for cool
 stars with molecular lines.

 In empirical libraries, the stars are
 assigned a spectral class based upon the appearance of spectral features 
 and therefore  are model
 independent.  Earlier reference libraries  did not have
 the required uniform range in 
atmospheric parameters. The empirical libraries assembled by \cite{ja84},
 \cite{pi85}, and \cite{si92} mostly
 contained solar-metallicity objects; the last two libraries also have lower resolution.
  The libraries assembled by  \cite{wo94} and \cite{ki91} had lower
 resolution, and the library by \cite{se96}
 had insufficient spectral coverage. At the inception of our 
 program (more than a decade ago) the database of stellar libraries
 was not satisfactory (particularly for metallicity coverage),
 hence we chose to develop our own 
 reference library. The situation has changed considerably now. 
 In the past decade, several new empirical libraries providing good
 spectral coverage at good resolution (R $\sim$ 2000 or better) have
 been developed.  For the optical region, libraries such as 
 STELIB (\cite{le03}), ELODIE
 (\cite{pr04}, \cite{pr07}), INDO-US (\cite{va04}),
  and more recently MILES (\cite{sa06}) have been provided while NGSL (\cite{gre06}),
  IRTF-Spex (\cite{ra09}), and XSL (\cite{ch11}) provide extended coverage
  from the ultraviolet to the infrared.
 With the help of softwares such as ULySS (\cite{ko09}) 
 large samples of stars can be classified and parametrized (see e.g.
 \cite{pr11}).  These empirical
 libraries are very important tools for building population-synthesis models
 and also for the automated classification and parametrization of stars.
 Notwithstanding its modest size, the reference library developed 
 by us is very useful for the present problem because of its uniform coverage 
 in metallicity, temperature, and gravity.

 From the medium-resolution spectra 
 metal-poor stars have been detected using different approaches;
 some are based upon the usage of strong features such as the Ca II lines
 (e.g. \cite{al00} on INT spectra),
 while others employ PCA or even full spectra
 (e.g. \cite{sn01}). A good account of stellar 
 parametrization approaches developed for handling data from different
 surveys can be found in the volume edited by \cite{ba08}.

 In this paper we used the artificial neural network (ANN) to estimate stellar parameters
  \te, log~$g$, and  \feh  and M$_{V}$  for a modest sample of candidate metal-poor
 stars using medium-resolution spectra.


\onlfig{1}{
\begin{figure*}
\centering
\includegraphics[angle=0,height=12cm,width=16cm]{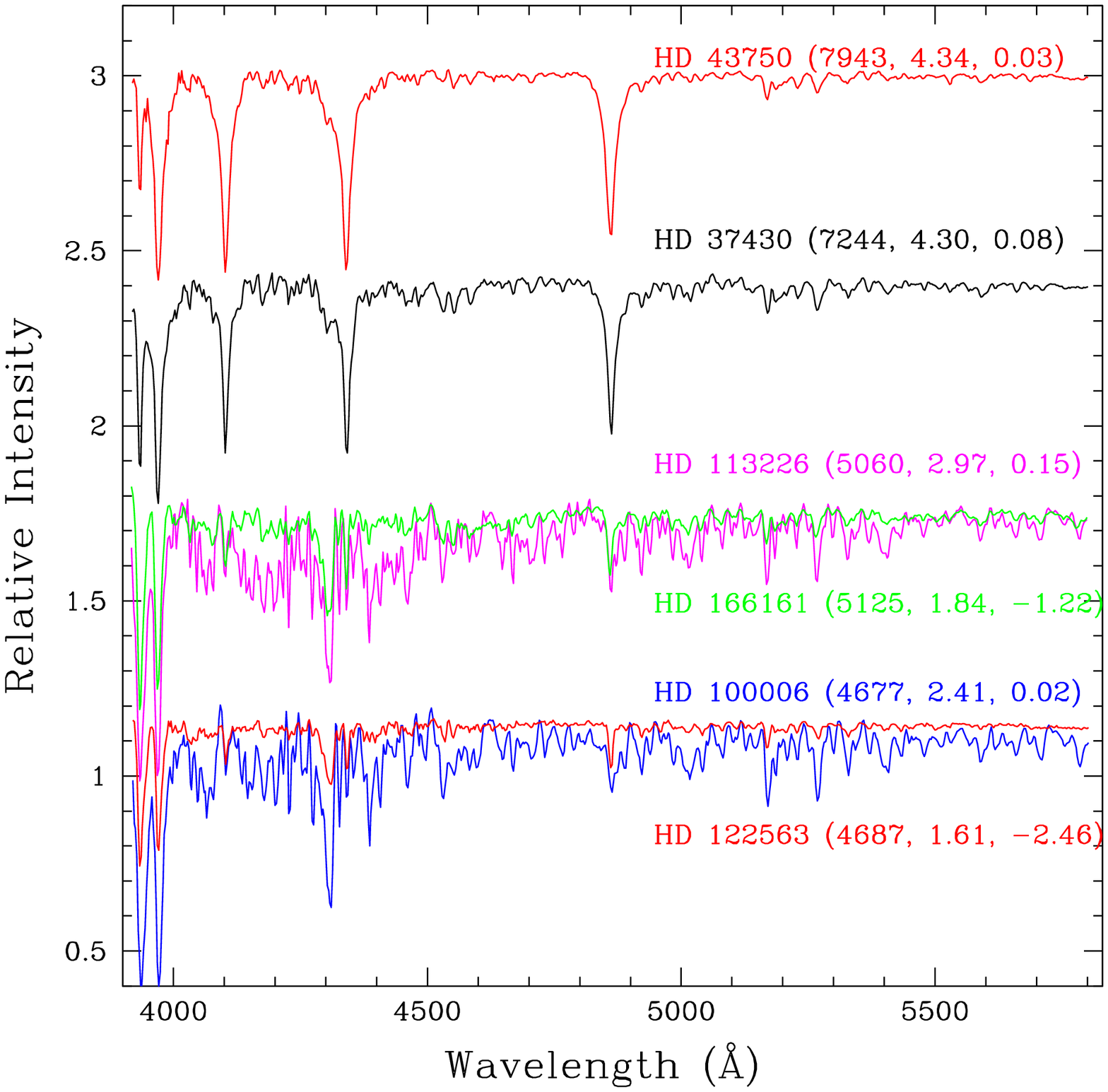}\\
\label{Figure 1}%
\caption{ Spectra of a selected sample of
stars with near-solar metallicity displayed in decreasing
temperature sequence from top to bottom.
The  spectra of a few  metal-poor stars are superposed on solar-metallicity
 stars of similar temperatures. The atmospheric parameters \te, log $g$, and \feh for
 each star are given in parenthesis.}
\end{figure*}}

\begin{figure*}
\centering
\includegraphics[angle=0,height=12cm,width=12cm]{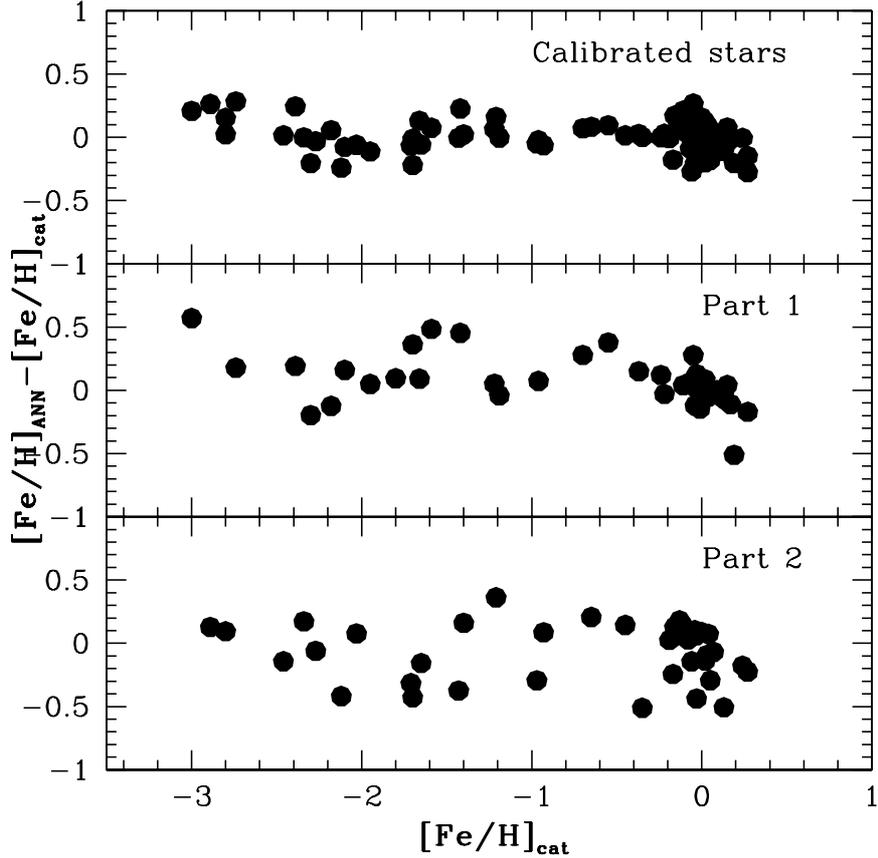}\\
\label{aop}%
\caption{Plot of [Fe/H]$_{ANN}$ $-$ [Fe/H]$_{Cat}$ 
 versus their catalog values, [Fe/H]$_{Cat}$ for all 
calibrated stars (top panel). The middle panel shows the result  for part 1
obtained using weights from the ANN trained for part 2. In the bottom panel  the weights 
from part~2 are applied to part~1. The rms error for the full sample (top panel) is 0.15 dex,
 for part 1 (middle panel) it is 0.31 dex, and for part 2 (bottom panel) it is 0.22 dex.}
\end{figure*}

\onlfig{3}{
\begin{figure*}
\centering
\includegraphics[bb=19 143 570 423]{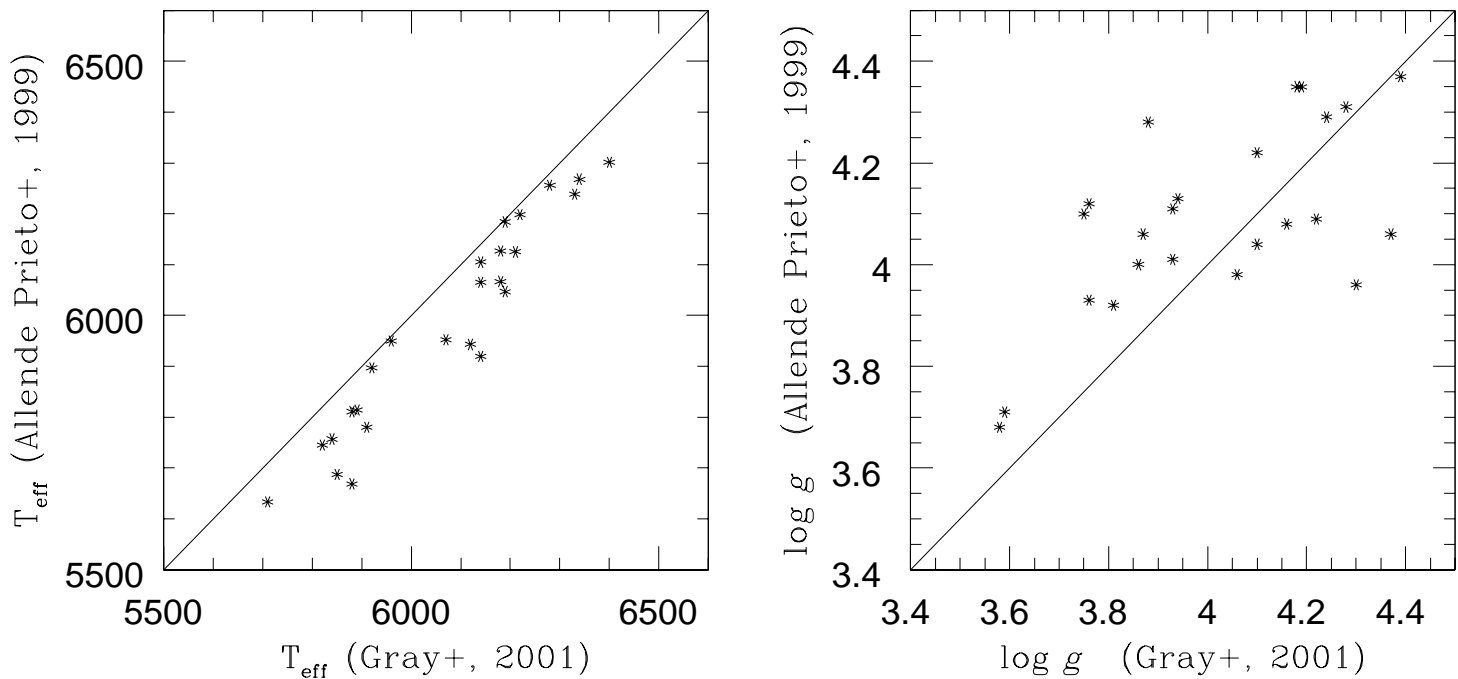}
\caption{ Comparison of  $T_{eff}$  and  log\,$g$ values for stars common  between
 Allende Prieto and Lambert (1999) and Gray, Graham and Hoyt (2001)
}
\label{Figure 3}
\end{figure*}}


\onltab{1}{
\onllongtab{3}{

{\footnotesize
\centering
\begin{center}
{\bf Table 1: List of observed stars and their parameters }\\
\end{center}
\begin{longtable}{clcccccccccccccc}
\hline\hline
\\
Sl No&Star   &   HIP& Vmag  & (B-V) &  \multicolumn{4}{c}{Literature} & \multicolumn{4}{c}{ANN} &Ref
\\
\hline\hline

&&&&&log\,g&  Teff&[Fe/H]&M$_{v}$ & log\,g&  Teff&[Fe/H]&M$_{v}$ \\
\\
\hline
\\
1	&    	HD~344		&   	655	& 	5.67	& 	1.119	& 	2.41	& 	4570.9	&      		& 	0.7     &	2.28	&	4635.6	&	$-$0.04	&	0.8&\\
2	& 	HD~496$^*$   	& 	765	& 	3.88	& 	1.013	& 	2.47	& 	4786.3	& 	$+$0.13	& 	0.7	&	2.58	&	4786.4	&$+$0.02&		0.7&		4     \\
3	& 	HD~587$^*$   	&  	840	& 	5.85	& 	0.973	& 	3.05	& 	4786.3	& 	$-$0.24	&	2.1 &		2.96	&	4812.4	&	$-$0.24	&	2.1&		1 \\
4	& 	HD~1529  	&  	1565	& 	7.95	& 	0.818	& 	3.70	& 	5248.1	&      		&     &			3.99	&	5294.8	&	$-$0.17&&\\
5	& 	HD~10142 	&  	7643	& 	5.94	& 	1.045	& 	2.68	& 	4786.3	&      		&  	0.9    &	2.40	&	4767.6	&	$-$0.34	&	0.9&&\\
6	& 	HD~14679 	&	10973 	& 	9.28	& 	0.652	& 	4.50	& 	5754.4	&      		&    &4.56&5754.8&$-$0.68&& \\
7	& 	HD~18709 	&	13902 	& 	7.39	& 	0.590	& 	4.40	& 	6025.6	&      		&   4.4  &4.30&6050.0&$-$0.50&4.4&\\
8	& 	HD~19445 	&	14594 	& 	8.05	& 	0.46 	& 	4.38	& 	6020.0  	&	$-$1.95 &5.1&4.34&5964.0&$-$2.06&4.6&	1   \\
9	& 	HD~19659 	&	14613 	& 	7.11	& 	0.684	& 	3.58	& 	5754.4	&      		& 2.3    &3.58&5664.4&$-$0.37&2.3&\\
10	& 	HD~20902$^*$ 	&	15863 	& 	1.82	&  	0.48 	& 	0.90	& 	6300.0  	& 	0.15 	&$-$4.5	&1.56&6316.4&$+$0.23&$-$4.2&	11    \\
11	& 	HD~21718 	& 	16270	& 	8.96	& 	1.163	& 	3.60	& 	4786.9	&      	&      &3.76&4695.6&$-$0.34&&\\
12	& 	HD~21925 	&	16479 	& 	8.30	& 	0.418	& 	4.42	& 	6606.9	&      	&    &4.31&6651.2&$-$0.12&& \\
13	& 	HD~22484$^*$ 	&	16852 	& 	4.28	& 	0.57 	& 	4.15	& 	5981.0  	&	$-$0.11 &3.6&4.14&6121.2&$-$0.05&3.2&	1   \\		
14	& 	HD~23190 	&	17575 	& 	6.83	& 	0.210	& 	4.20	& 	7943.3	&      	& 2.1     &4.35&7848.0&0.21&2.4&\\
15	& 	HD~23650 	&	17887 	& 	9.01	& 	0.582	& 	4.55	& 	6025.6	&      	& 5.0    &4.54&6008.8&$-$0.23&4.6&\\
16	& 	HD~26519 	&	19501 	& 	7.86	& 	0.440	& 	4.42	& 	6606.9	&      	& 3.9  &4.49&6555.6&$-$0.48&3.8 & \\
17	& 	HD~26749 	&	19767 	& 	6.74	& 	0.677	& 	4.11	& 	5754.4	&      	& 4.0  &4.43&5543.2&$-$0.60&4.3&  \\
18	& 	HD~27045 	&	19990 	& 	4.93	& 	0.259	& 	4.30	& 	7585.8	&      	& 2.6  &4.25&7654.8&0.17&2.6 &  \\
19	& 	HD~27174 	& 	20334	& 	8.25	& 	1.071	& 	3.43	& 	4677.4	&      	& 3.7   &3.36&4634.0&$-$0.05&3.8 & \\
20	& 	HD~29140$^*$ 	&	21402 	& 	4.25	& 	0.184	& 	3.81	& 	7943.3	&	0.27 & 0.9	&3.94&7835.2&$-$0.006&1.2&	17    \\
21	& 	HD~30177 	& 	21850	& 	8.41	& 	0.773	& 	4.30	& 	5495.4	&      	&  4.7 &4.89&5483.0&$-$0.27&5.2 & \\
22	& 	HD~284908	& 	22684	& 	9.28	& 	1.128	& 	3.73	& 	4677.4	&      	&   &4.10&5369.0&$-$0.38&&   \\
23	& 	HD~31109 	&	22701 	& 	4.36	& 	0.257	& 	3.40	& 	7244.4	&      	&   0.1&3.40&7362.0&0.18& 0.1&  \\
24	& 	HD~32890 	& 	23668	& 	5.71	& 	1.166	& 	2.70	& 	4570.9	&      	&&3.089    &4717.2&$-$0.109&&  \\
25	& 	HD~33111 	&	23875 	& 	2.78	& 	0.161	& 	3.70	& 	7943.3	&      	& 0.6   &3.85&7850.4&0.21&0.3&  \\
26	& 	HD~33419 	& 	24041	& 	6.11	& 	1.098	& 	2.50	& 	4570.9	&      	&  1.2   &2.29&4655.2&0.07&1.2 &\\
27	& 	HD~34303 	& 	24665	& 	6.85	& 	1.061	& 	2.85	& 	4677.4	&      	&  2.2   &2.87&4679.2&$-$0.03&2.2 &\\
28	& 	HD~34500 	&	24730 	& 	7.41	& 	0.204	& 	4.36	& 	7943.3	&      	&  2.8  &4.16&7889.2&0.22&2.7 & \\
29	& 	HD~36079$^*$ 	& 	25606	& 	2.81	& 	0.807	& 	2.54	& 	5248.1	&	0.05 &	&2.34&5275.6&$-$0.13&&	1   \\
30	& 	HD~36153 	&	25651 	& 	7.32	& 	0.305	& 	4.28	& 	7244.4	&      	&  2.8 &4.07&7360.4&0.12& 2.7 & \\
31	& 	HD~36673$^*$ 	&	25985 	& 	2.59	&  	0.21 	& 	1.10	& 	7400.0 	&	0.04 & $-$5.4 &1.25&7357.6&0.01&$-$5.7	&14     \\
32	& 	HD~37192 	& 	26219	& 	5.76	& 	1.120	& 	2.40	& 	4570.9	&      	& 0.8     &2.30&4558.4&$-$0.05&0.7&\\
33	& 	HD~37430 	&	26412 	& 	6.15	& 	0.322	& 	4.30	& 	7244.4	&      	&  2.9  &4.28&7283.6&0.08&2.7 & \\
34	& 	HD~37984$^*$ 	& 	26885	& 	4.90	& 	1.144	& 	2.21	& 	4570.9	&	$-$0.55& 0.07&2.21&4747.6&$-$0.45&0.3&	1  \\
35	& 	HD~37613 	&	26996 	& 	7.84	& 	0.455	& 	4.20	& 	6606.9	&      	& 3.0  &4.36&6565.6&$-$0.02&2.9 & \\
36	& 	SAO~58437	&	27361 	& 	9.19	& 	0.372	& 	4.40	& 	6918.3	&      	& 4.3   &4.57&6939.6&$-$0.37&4.0 & \\
37	& 	HD~39425$^*$ 	& 	27628	& 	3.12	& 	1.146	& 	2.31	& 	4570.9	&	$+$0.13&1.0 	&2.29&4572.8&$+$0.081&1.0&	1  \\
38	& 	HD~41393 	&	28654 	& 	6.88	& 	0.201	& 	4.29	& 	7943.3	&      	& 2.3  &4.13&7874.8&0.26&2.3 &  \\
39	& 	BD+191185$^*$ 	&	28671 	& 	9.31	&  	0.588 	& 	4.29	& 	5440.0  	&	$-$1.21& &4.74&5571.2&$-$1.05&&	   \\
40	& 	HD~41116$^*$ 	& 	28734	& 	4.16	& 	0.835	& 	2.97	& 	5248.1	&	$-$0.01 &0.8&3.109&5323.2&$-$0.045&1.0&	1  \\
41	& 	HD~41547 	&	28854 	& 	5.88	& 	0.374	& 	3.90	& 	6918.3	&	$-$0.10 & &4.054&7069.6&0.095&&	12   \\
42	& 	HD~41712$^*$ 	&	29002 	& 	6.94	& 	0.455	& 	3.90	& 	6606.9	&	$-$0.03 &2.3 &4.10&6535.2&$+$0.06&2.4&	12  \\
43	& 	HD~44007$^*$	&	29992 	& 	8.06	&  	0.84 	& 	2.00	& 	4830.0  	&	$-$1.71 &&2.03&4845.6&$-$1.77&&	1  \\
44	& 	HD~43750 	&	30165 	& 	7.44	& 	0.201	& 	4.34	& 	7943.3	&      	&  2.8 &4.28&7768.0&0.03&2.7 &  \\
45	& 	HD~43771 	&	30275 	& 	7.43	& 	0.209	& 	4.33	& 	7943.3	&      	&  2.6 &4.24&7822.0&0.15&2.2 &  \\
46	& 	HD~46355 	& 	30932	& 	5.20	& 	1.087	& 	2.26	& 	4677.4	&      	&  0.3 &2.25&4679.2&$-$1.21& 0.3&  \\
47	& 	HD~48329$^*$ 	&	32246 	& 	3.02	&  	1.40 	& 	0.80	& 	4582.0  	&	$-$0.05 &$-$4.2 &1.32&4511.6&$+$0.21&$-$4.1&	1  \\
48	& 	CD-333337	&	33221 	& 	9.03	&  	0.48 	& 	4.11	& 	5930.0 	&	$-$1.40 & &4.146&5973.6&$-$1.74&&	3  \\
49	& 	HD~52622 	&	33577 	& 	6.46	& 	0.389	& 	3.68	& 	6918.3	&      	&1.5   &2.963&6901.2&0.181&1.3 & \\
50	& 	HD~56935 	& 	35154	& 	7.69	& 	0.653	& 	3.75	& 	4786.9	&      	& 4.0   &3.77&4781.2&$-$0.07&4.0&  \\
51	& 	HD~56221 	&	35341 	& 	5.87	& 	0.181	& 	3.94	& 	7943.3	&      	&  1.3 &3.90&7855.6&0.17&1.2 &  \\
52	& 	HD~58431$^*$ 	&	36059 	& 	7.84	& 	0.331	& 	4.31	& 	7244.4	&	$-$0.07 &3.0 &4.32&7393.2&$-$0.16&2.9	&12   \\
53	& 	HD~58946$^*$ 	&	36366 	& 	4.16	& 	0.31 	& 	4.47	& 	7145.0  	&	$-$0.17 &2.8&4.11&7082.4&$-$0.35&2.9&	12   \\
54	& 	HD~61295$^*$ 	&	37339 	& 	6.16	& 	0.374	& 	3.70	& 	6918.3	&	0.02  &1.5 &3.47&6939.2&0.14&1.4&	1   \\
55	& 	HD~62781 	&	37710 	& 	5.80	& 	0.320	& 	4.16	& 	7244.4	&      & 2.6   &4.17&7306.4&0.09&2.4&  \\
56	& 	HD~62345$^*$ 	&	37740 	& 	3.57	& 	0.93 	& 	2.90	& 	5000.0  	&	$-$0.16&0.4&2.54&4892.0&+0.01&0.3&	1  \\
57	& 	HD~62196 	&	37802 	& 	7.67	& 	0.313	& 	4.37	& 	7244.4	&      & 3.6   &4.518&6952.8&$-$0.78&3.6 & \\
58	& 	HD~62509$^*$ 	&	37826 	& 	1.15	& 	1.00 	& 	2.75	& 	4865.0  	&	$-$0.04& 1.0&2.78&4813.2&$+$0.04&1.0&	1  \\
59	& 	HD~63660 	& 	38146	& 	5.32	& 	0.751	& 	3.02	& 	5495.4	&      & 0.3  &2.93&5479.2&0.10&0.2 & \\
60	& 	HD~63700$^*$ 	&	38170 	& 	3.34	&  	1.25 	& 	1.15	& 	4990.0  	&	0.24 &	&0.825&4684.8&$+$0.317&	&1  \\
61	& 	HD~63791$^*$ 	&	38621 	& 	7.92	&  	      & 		1.80	& 	4750.0  	&	$-$1.65 &&1.79&4761.6&$-$1.70&	&1  \\
62	& 	HD~65228 	&	38835 	& 	4.20	&  	0.73 	& 	1.70	& 	5900.0  	& 	0.52 &  &1.663&5728.0&0.138&&	1   \\
63	& 	HD~67078 	&	39565 	& 	6.62	& 	0.448	& 	3.81	& 	6606.9	&      & 2.0   &4.09&6583.2&0.01&2.3 &\\
64	& 	HD~65871 	&	39616 	& 	8.16	& 	0.529	& 	4.40	& 	6309.6	&      &  4.3  &4.42&6272.8&$-$0.47&4.0& \\
65	& 	HD~70110$^*$ 	&	40858 	& 	6.18	& 	0.607	& 	4.01	& 	6025.6	&	0.07&3.1	&4.29&5994.8&$-$0.01&3.3&	1  \\
66	& 	HD~69960 	& 	41022	& 	8.00	& 	0.756	& 	4.06	& 	5495.4	&      & 4.0 &3.86&5455.2&$-$0.02&3.8 &  \\
67	& 	HD~71973 	&	42249 	& 	6.31	& 	0.308	& 	3.85	& 	7244.4	&      & 1.7   &3.59&7183.2&$-$0.04&1.5 & \\
68	& 	HD~73764 	& 	42528	& 	6.60	& 	0.899	& 	3.22	& 	5011.9	&      &  2.0  &3.08&4992.4&$-$0.11&2.0 & \\
69	& 	HD~74706 	&	42928 	& 	6.10	& 	0.195	& 	4.11	& 	7943.3	&      & 1.6  &4.05&7824.8&0.19&1.5&   \\
70	& 	HD~76218 	&	43852 	& 	7.69	& 	0.771	& 	4.59	& 	5495.4	&      & 5.6   &4.47&5426.0&$-$0.19&5.2& \\
71	& 	HD~76582 	&	44001 	& 	5.68	& 	0.209	& 	4.25	& 	7943.3	&      & 2.2   &4.04&7792.0&0.20&2.3&  \\
72	& 	HD 76932$^*$ 	&	44075 	& 	5.86	& 	0.53 	& 	4.37	& 	5965.0  	&	$-$0.82& &4.128&5896.8&$-$1.315&&	9    \\
73	& 	HD~76617 	&	44103 	& 	8.17	& 	0.596	& 	4.12	& 	6025.6	&      & 3.4  &4.08&6005.6&$-$0.04&3.2 & \\
74	& 	HD~76909 	& 	44137	& 	7.84	& 	0.756	& 	4.22	& 	5495.4	&      & 4.4 &4.16&5528.8&0.06&4.3&   \\
75	& 	HD~78752 	&	44915 	& 	7.84	& 	0.602	& 	4.01	& 	6025.6	&      & 3.2 &3.96&5935.2&$-$0.24&3.5&   \\
76	& 	HD~233608	& 	45098	& 	9.40	& 	0.879	& 	4.34	& 	5248.1	&      &  &4.48&5362.4&$-$0.08&&   \\
77	& 	HD~76990 	&	45421 	& 	6.30	& 	0.339	& 	3.72	& 	6918.3	&      & 1.5  &3.63&6993.6&$-$0.18&1.6&   \\
78	& 	HD 83212$^*$ 	&	47139 	& 	8.34	&  	1.09 	& 	1.00	& 	4763.0  	&	$-$1.47& &2.662&4932.8&$-$1.37&&	2   \\
79	& 	HD~83808 	&	47508 	& 	3.52	& 	0.516	& 	3.23	& 	6309.6	&      & 0.4   &3.42&6390.8&$-$0.17&0.4& \\
80	& 	HD~84441$^*$ 	&	47908 	& 	2.97	& 	0.81 	& 	1.70	& 	5300.0  	&	0.17  &$-$1.4	&2.611&6014.4&0.048&$-$1.4&	1   \\
81	& 	HD~84850 	&	47913 	& 	6.22	& 	0.461	& 	3.70	& 	6606.9	&      & 1.7  &3.92&6532.4&0.08&1.7 & \\
82	& 	HD~84937$^*$ 	&	48152 	& 	8.28	& 	0.41 	& 	4.00	& 	6211.0  	&	$-$2.34 &3.7&4.439&6614.8&$-$2.18&3.7&	1   \\
83	& 	G43-5$^*$    	&      		&	12.52	&       0.65 	& 	4.66	& 	5310.0  	&	$-$2.12 &  &4.71&5338.8&$-$2.36&&	3  \\
84	& 	HD~85379 	& 	48347	& 	7.34	& 	1.187	& 	3.20	& 	4570.9	&      & 3.0    &3.19&4540.4&$-$0.01&3.0 &\\
85	& 	HD~85444$^*$ 	& 	48356	& 	4.11	& 	0.918	& 	2.48	& 	5011.9	&	$-$0.14&$-$0.5&2.61&5109.6&$-$0.02&$-$0.5	&1  \\
86	& 	HD~085773	&	48516 	& 	9.43	&  	1.16 	& 	0.99	& 	4470.0  	&	$-$2.27 & &2.318&4659.2&$-$2.08&&	3 \\
87	& 	HD~85844 	&	48590 	& 	8.23	& 	0.263	& 	4.37	& 	7585.8	&      &  3.4  &4.38&7554.8&$-$0.14&3.5&  \\
88	& 	HD~87427 	&	49339 	& 	5.70	& 	0.303	& 	3.70	& 	7244.4	&      & 1.2  &3.70&7355.2&0.16&1.2 &  \\
89	& 	HD 87140$^*$ 	&	49371 	& 	9.00	&  	0.70 	& 	2.58	& 	4940.0  	&	$-$2.02& &2.573&5086.0&$-$1.85&&	5   \\
90	& 	G43-33$^*$ 	&	49988 	& 	7.85	&  	0.55 	& 	4.30	& 	5925.0 	&	$-$0.37& &4.17&5948.0&$-$0.344&&	   \\
91	& 	G54-21$^*$ 	&	50355 	& 	7.62	&  	0.60 	& 	4.48	& 	5862.0  	&	$-$0.03& &4.60&5839.0&$-$0.243&&	   \\
92	& 	HD~89086 	&	50364 	& 	7.62	& 	0.468	& 	4.22	& 	6606.9	&      & 3.2  &4.29&6616.0&$-$0.10&3.0&  \\
93	& 	HD~89449$^*$ 	&	50564 	& 	4.80	& 	0.44 	& 	4.14	& 	6385.0  	&	0.09 &3.2	&4.26&6584.4&0.09&3.1&	1    \\
94	& 	HD~89962 	& 	50851	& 	6.06	& 	1.119	& 	2.90	& 	4677.4	&      & 1.8   &2.85&4679.6&$-$2.01&1.8&  \\
95	& 	HD~90860 	&	51414 	& 	7.01	& 	0.622	& 	3.74	& 	6025.6	&      & 2.2 &3.65&5978.4&$-$0.56&2.6&   \\
96	& 	HD~91135 	&	51475 	& 	6.51	& 	0.534	& 	3.60	& 	6309.6	&      &1.6  &3.45&6244.8&0.08&1.6&   \\
97	& 	HD~91669 	& 	51789	& 	9.70	& 	0.877	& 	4.40	& 	5248.1	&      &  &4.43&5182.4&0.09& &  \\
98	& 	HD~91948$^*$ 	&	52064 	& 	6.77	& 	0.465	& 	3.99	& 	6606.9	&	$-$0.03 &2.5 &4.09&6449.2&$-$0.001&2.5&	12  \\
99	& 	G58-23$^*$  	&	52958 	& 	9.96	& 	0.60 	& 	4.40	& 	5540.0  	&	$-$0.97 & 5.2&4.37&5464.0&$-$1.02&5.2&	3  \\
100	& 	HD~94028$^*$ 	&	53070 	& 	8.21	& 	0.498	& 	4.20	& 	5900.6	&	$-$1.55 &&4.223&6062.8&$-$1.80&&	1 \\
101	& 	BD-163141	& 	53174	& 	10.4	& 	0.906	& 	4.21	& 	5011.9	&      &   &4.507&4812.8&$-$0.342& & \\
102	& 	HD~94771 	& 	53437	& 	7.37	& 	0.752	& 	3.90	& 	5495.4	&      & 3.7  &3.85&5456.0&$-$0.04&3.4 & \\
103	& 	HD~95272$^*$ 	& 	53740	& 	4.08	& 	1.079	& 	2.34	& 	4677.4	&	$-$0.22&0.4 &2.09&4654.8&$-$0.19&0.3&	1  \\
104	& 	HD~95364 	&	53851 	& 	8.62	& 	0.690	& 	4.20	& 	5754.4	&      & 4.0  &4.08&5628.8&$-$0.40&4.0&  \\
105	& 	HD~95532 	&	53886 	& 	7.58	& 	0.543	& 	4.10	& 	6309.6	&      & 3.2  &4.24&6306.0&$-$0.09&3.3 & \\
106	& 	HD~96833$^*$ 	& 	54539	& 	3.00	& 	1.144	& 	2.08	& 	4570.9	&	$-$0.13&$-$0.2 &2.03&4524.0&$-$0.01&$-$0.2&	1  \\
107	& 	HD~97336 	&	54741 	& 	8.15	& 	0.357	& 	4.35	& 	6918.3	&      &  3.5  &4.37&6698.0&$-$0.76&3.6&  \\
108	& 	HD~97998 	&	55013 	& 	7.36	& 	0.626	& 	4.57	& 	5754.4	&      & 5.2  &4.45&5798.0&$-$0.42&4.7 & \\
109	& 	HD~98175 	&	55126 	& 	6.85	& 	0.328	& 	4.05	& 	7244.4	&      & 2.0  &3.98&7127.6&$-$0.10&2.2&   \\
110	& 	HD~98579 	& 	55374	& 	6.68	& 	1.124	& 	2.84	& 	4570.9	&      &  1.8 &2.69&4640.4&$-$0.31&1.9&   \\
111	& 	HD~100006$^*$ 	& 	56146	& 	5.54	& 	1.056	& 	2.41	& 	4677.4	&	+0.02&0.5 	&2.42&4702.4&$-$0.18&0.3&	1  \\
112	& 	HD~101165	&	56795 	& 	9.18	& 	0.615	& 	4.34	& 	6025.6	&      & 4.2  &4.20&6005.2&$-$0.29&3.7 & \\
113	& 	HD~101501$^*$ 	&	56997 	& 	5.32	&	+0.710	& 	4.69	& 	5538.0  	&	0.03 & 5.4	&4.58&5444.4&$-$0.08&5.0&	1  \\
114	& 	HD~102070$^*$ 	&	57283 	& 	4.72	&  	0.97 	& 	2.57	& 	4870.0  	&	$-$0.11 &$-$0.4 &2.28&4879.2&$+$0.10&$-$0.5&	1 \\
115	& 	HD~102902	&	57759 	& 	7.36	& 	0.701	& 	3.81	& 	5754.4	&      & 2.6   &4.01&5851.6&$-$0.25&3.1& \\
116	& 	HD~103095$^*$ 	&	57939 	& 	6.45	&	+0.75 	& 	4.50	& 	5000.0  	&	$-$1.59 & &4.72&5020.0&$-$1.52&&	1  \\
117	& 	HD~104163	& 	58502	& 	8.48	& 	0.879	& 	3.68	& 	5011.9	&      &   &3.67&5130.8&$-$0.39&&  \\
118	& 	HD~107325	& 	60170	& 	5.52	& 	1.091	& 	3.04	& 	4677.4	&      & 2.1   &3.01&4767.2&$-$0.16&2.1&  \\
119	& 	HD~107610	& 	60305	& 	6.33	& 	1.115	& 	2.61	& 	4570.9	&      &   1.4   &2.47&4611.2&$-$0.10&1.4&\\
120	& 	HD~107700$^*$ 	&	60351 	& 	4.78	& 	0.515	& 	3.14	& 	6309.6	&	$-$0.06 &0.2 &3.25&6498.4&$-$0.33&0.2&	12  \\
121	& 	HD~107752$^*$ 	&	60387 	&	10.07	&  	0.75 	& 	2.07	& 	4710.0  	&	$-$2.74 &&2.18&4760.4&$-$2.45&&	3 \\
122	& 	HD~108317$^*$ 	&	60719 	& 	8.04	&  	      & 		3.33	& 	5310.0  	&	$-$2.27 &1.3&3.21 &5186.0&$-$2.30&1.5&	3  \\
123	& 	G13-38$^*$  	&	60747 	&	10.51	& 	0.71 	& 	4.60	& 	5220.0  	&	$-$0.96 &5.7 &4.61&5134.0&$-$0.98&5.4&	3 \\
124	& 	HD~108506	&	60813 	& 	6.23	& 	0.430	& 	3.64	& 	6606.9	&      & 1.4  &3.84&6636.8&$-$0.02&1.3 & \\
125	& 	HD~109358$^*$ 	&	61317 	& 	4.26	&	$+$0.59 	& 	4.52	& 	5879.0  	&	$-$0.19 &4.6 &4.45&5971.6&$-$0.19&4.6&	1   \\
126	& 	HD~109379$^*$ 	&	61359 	& 	2.65	& 	$+$0.89 	& 	2.20	& 	5125.0  	& 	0.27 &$-$0.5	&2.38&5150.0&0.12&$-$0.4&	1   \\
127	& 	G59-27$^*$	&	61545 	& 	10.86	& 	$+$0.425	& 	3.50	& 	6150.0  	&	$-$2.20 &	&4.173&6072.0&$-$2.27&& 19   \\
128	& 	HD~110317J$^*$ 	&	61910	& 	5.17	& 	0.432	& 	3.34	& 	6606.9	&	0.00 &0.5	&3.57&6626.4&0.15&0.4	&1  \\
129	& 	HD~110646	& 	62103	& 	5.91	& 	0.850	& 	3.23	& 	5248.1	&      & 1.7  &3.35&5191.6&$-$0.19&2.0 & \\
130	& 	G60-46$^*$    	&      &		11.00	&      &       		4.59	& 	5300.0  	&	$-$1.19 & &4.58&5289.2&$-$1.19&&	3  \\
131	& 	HD~113226$^*$ 	&	63608 	& 	2.83	& 	$+$0.94 	& 	2.97	& 	5060.0  	& 	0.15 &	&2.908&5063.2&0.025&&	1    \\
132	& 	HD~114435	&	64332 	& 	5.78	& 	0.521	& 	3.34	& 	6309.6	&      & 0.9 &2.865&6673.6&0.12&0.7 &  \\
133	& 	HD~115772$^*$ 	&	65047 	& 	9.63	&  	0.84 	& 	2.56	& 	4930.0  	&	$-$0.70 & &2.48&4933.6&$-$0.63&&	3 \\
134	& 	HD~118253	& 	66381	& 	7.58	& 	0.875	& 	3.47	& 	5011.9	&      & 2.9   &3.44&5116.8&$-$0.56&2.9& \\
135	& 	HD~121370$^*$ 	&	67927 	& 	2.68	& 	0.59 	& 	3.83	& 	6068.0  	& 	0.19 &	2.4&3.73&5943.6&$-$0.02&2.3&	1    \\
136	& 	HD~122167	&	68367 	& 	8.67	& 	0.570	& 	4.41	& 	6025.6	&      &   &4.20&5906.4&$-$0.36&&  \\
137	& 	HD~121930	& 	68375	& 	7.58	& 	1.199	& 	3.10	& 	4570.9	&      & 2.7  &3.03&4628.4&$-$0.23&2.7 &  \\
138	& 	G64-37$^*$ 	&	68592 	& 	11.149	&  	0.359 	& 	4.20	& 	6377.0 	&	$-$3.0& &4.22&6477.0&$-$2.792&&	   \\
139	& 	HD~122563$^*$ 	&	68594 	& 	6.20	&  	0.90 	& 	1.61	& 	4687.0  	&	$-$2.46 &$-$0.9&1.76&4668.0&$-$2.44&$-$1.0&	3 \\
140	& 	BD+092870$^*$ 	&	69746 	& 	9.45	&  	      & 		1.62	& 	4672.0  	&	$-$2.39 &1.2 &2.47&4865.0&$-$2.14&1.1&	3 \\
141	& 	HD 126053a$^*$ 	&	70319 	& 	6.30	&  	0.60 	& 	4.50	& 	5662.0  	&	$-$0.45&$+$5.07 &4.31&5683.2&$-$0.435&&	   \\
142	& 	HD~126354	&	70576 	& 	4.33	& 	0.434	& 	3.01	& 	6606.9	&      &  $-$0.6 &2.92&6524.8&$-$0.41&$-$0.6&  \\
143	& 	HD~127665 	&	71053 	& 	3.58	& 	1.29 	& 	2.22	& 	4260.0  	&	$-$0.17 &&2.194&4384.0&$-$0.039&&	1   \\
144	& 	HD~127739$^*$ 	&	71115 	& 	5.91	& 	0.391	& 	4.02	& 	6918.3	& 	0.08 & 2.3 &4.05&6980.8&0.06&2.1&	13 \\
145	& 	HD~129401	&	72041 	& 	8.68	& 	0.607	& 	4.26	& 	6025.6	&      & 3.8  &4.20&6003.2&$-$0.09&3.8 & \\
146	& 	HD~130169	&	72455 	& 	7.13	& 	0.521	& 	3.93	& 	6309.6	&      & 2.7  &4.15&6258.8&$-$0.22&3.2&  \\
147	& 	BD+452224	& 	72504	& 	10.7	& 	1.110	& 	3.96	& 	4570.9	&      &   &3.91&4549.2&$-$0.69& &  \\
148	& 	HD~132047	& 	73065	& 	7.66	& 	1.060	& 	3.38	& 	4677.4	&      &  3.5 &3.50&4731.6&$-$0.19&3.3 &  \\
149	& 	G99-40$^*$ 	&	 	& 	9.19	&  	0.61 	& 	4.08	& 	5970.0 	&	$-$0.35& &4.26&5994.0&$-$0.348&&	   \\
150	& 	HD~132475$^*$ 	&	73385 	& 	8.57	& 	0.59 	& 	3.76	& 	5550.0  	&	$-$1.70 & 3.7 &3.83&5594.4&$-$1.71&4.1&	3  \\
151	& 	HD~134440 	&	74234 	& 	9.44	& 	0.85 	& 	4.70	& 	4790.0  	&	$-$1.43 &&4.64&4834.4&$-$1.43&&	1  \\
152	& 	HD~136202$^*$ 	&	74975 	& 	5.10	& 	0.54 	& 	4.07	& 	6077.0  	&	$-$0.15 &&3.921&6223.6&$-$0.191&&	1  \\
153	& 	HD~147397	& 	80163	& 	8.35	& 	1.323	& 	3.81	& 	4786.9	&      &    &3.60&4681.6&$-$0.15&&  \\
154	& 	HD~148408	&	80630 	& 	9.62	&       		0.71 &		4.55.0 	& 	5200.0  	& 	$-$0.8 & &3.925&4933.2&$-$1.40&&	3 \\
155	& 	HD~149996$^*$ 	&	81461 	& 	8.49	& 	0.62 	& 	4.1 	& 	5600.0  	&	$-$0.65 &4.3&4.17&5566.8&$-$0.56&4.3&	1   \\
156	& 	HD~153210$^*$	&	83000 	& 	3.20	& 	1.16 	& 	2.62	& 	4560.0  	&	$-$0.13 &1.0 &2.455&4592.0&0.12&1.0&	1   \\
157	& 	BD+173248$^*$ 	&	85487 	& 	9.37	&  	0.66 	& 	2.94	& 	4995.0  	&	$-$2.03 & 2.2 &3.247&5170.0&$-$2.09&2.3&	3 \\
158	& 	HD~161096$^*$ 	&	86742 	& 	2.77	& 	1.16 	& 	1.70	& 	4475.0  	& 	0.00 &	&2.225&4507.2&0.061&&	1  \\
159	& 	HD~161797$^*$ 	&	86974 	& 	3.41	&	0.75 	& 	3.70	& 	5520.0  	& 	0.04 & 3.8	&3.99&5563.2&0.12&4.0&	1 \\
160	& 	HD~165195$^*$ 	&	88527 	& 	7.34	&  	1.29 	& 	1.45	& 	4507.0  	&	$-$2.18 &$-$0.9 &2.044&4724.0&$-$1.91&$-$0.8&	1 \\
161	& 	HD~166161$^*$ 	&	88977 	& 	8.16	&  	0.98 	& 	1.84	& 	5125.0  	&	$-$1.22 & 0.7&2.0&5148.4&$-$1.15&0.5&	3 \\
162	& 	G141-19$^*$  	&	90957 	&	10.55	&  	0.64 	& 	4.00	& 	5400.0  	&	$-$2.30 &&3.87&5396.4&$-$2.5&&	1  \\
163	& 	HD~185144 	&	96100 	& 	4.70	&	0.79 & 		4.40	& 	5143.0  	&	$-$0.25 &  &4.41&5588.0&$-$0.429&&	6    \\
164	& 	HD~188512$^*$ 	&	98036 	& 	3.71	& 	0.86 	& 	3.60	& 	5100.0  	&	$-$0.30 &&3.525&5017.2&$-$0.35&&	1   \\
165	& 	BD-185550$^*$ 	&	98339 	& 	9.35	&  	0.92 	& 	1.87	& 	4785.0  	&	$-$2.89 &0.7 &1.86&4783.6&$-$2.62&0.7&	3 \\
166	& 	CS22877-1$^*$ 	&	 	& 		&  	 	& 	1.00	& 	4500.0 	&	$-$2.80& &1.02&4512.0&$-$2.648&&	   \\
167	& 	CS22169-35$^*$ 	&	 	& 	12.9	&  	 	& 	1.50	& 	5000.0 	&	$-$2.80& &1.38&5017.0&$-$2.778&&	   \\
\hline
\end{longtable}
}
$^*$ indicates stars with known metallicity, the references for metallicity are given below. \\
1. \cite{ca01};  2. \cite{ry95}; 3. \cite{sn01};
4. \cite{gr86}; 5. \cite{to92}; 6.  \cite{oi74};
7. \cite{ax94}; 8.  \cite{gr01},\\ 9.  \cite{ed93};
10. \cite{lu81}; 11. \cite{lu85}; 12. \cite{mo04};
13. \cite{ba90}; 14.\cite{ve95}; 15. \cite{bu89};\\
16. \cite{ad94}; 17. \cite{pa73}; 
18. \cite{to99}; 19. \cite{sp94}.
}}

\noindent
In section 2 we describe the stellar  spectra database developed by us and the 
subset used for calibrating the ANN. Section 3 describes the observations
and spectral analysis.  Section 4 deals with the network configuration
and the adopted network-training approach. We present in section 5 the 
atmospheric parameters and calibration errors and the use of trained 
networks to estimate the parameters for a sample of candidate metal-poor 
stars and some unexplored  field stars. The determination of absolute magnitudes
is presented in section 6, derived parameters for candidate metal-poor 
 stars are given in section 7. We summarize our results in section 8.

\onltab{2}{
\onllongtab{3}{
{\footnotesize
\centering
\begin{center}
{\bf Table 2: Estimated atmospheric parameters for candidate metal-poor stars }\\
\end{center}
\begin{longtable}{lcccccc}
\hline\hline
\\
 Star    &     log\,g  & [Fe/H] &  T$_{eff}$ & (B-V) & T$_{eff}$(ANN)$-$ &M$_{v}$  \\
         &  [ANN]  &    [ANN]   &  [ANN]     &       & T$_{eff}$[B-V] &  \\
\\
\hline
\\
EC~00451-2737 &   4.4  &  $-$1.97  & 6278.4  &  ---- & ----  &   4.1  \\
EC~01374-3243 &   4.5  &  $-$1.35  & 5890.0  &   ----&  ---- &    5.4  \\
EC~03531-5111 &    3.3  &  $-$0.43  & 7164.8  &  ---- & ----  &   4.1   \\
EC~04555-1409 &    3.3  &  $-$0.32  &  5862.0  &   ----&  ---- &   $-$0.3  \\
EC~05148-2731 &   4.7  &  $-$0.58  & 7934.0  &   ----&   ----&     5.6*   \\
EC~09523-1259 &   4.4  &  $-$1.37  & 6572.4  &   $+$0.47& 159.4 &     5.4*  \\
EC~10004-1405a&    3.9  &  $-$1.19  & 5864.0  &   ----&  ---- &    3.5   \\
EC~10004-1405b&    3.9  &  $-$1.02  & 5702.0  &   ----&  ---- &    4.2   \\
EC~10262-1217 &    4.6  &  $-$1.16  & 6636.0  &  $+$0.38 & $-$187  &   4.5   \\
EC~10292-0956 &    4.5  &  $-$0.46  &  6260.4  &  $+$0.58 & 291.4  &   4.4  \\
EC~10488-1244 &    4.1  &  $-$0.33  &  6507.2  &  $+$0.50 & 222.2  &   2.7   \\
EC~11091-3239 &    4.4  &   0.09  &  5957.2  &  $+$0.54 &  $-$164.8 &    3.7  \\
EC~11175-3214 &   4.7  &  $-$1.40  & 7702.4  &   $+$0.43&   1112.4&     5.6*   \\
EC~11260-2413 &   4.7  &  $-$0.96  & 7523.2  &   $+$0.39&   748.2&     5.4*   \\
EC~11553-2731a&    4.4  &  $-$0.44  &  6374.8  &  ---- & ----  &   3.7   \\
EC~11553-2731b&    4.2  &  $-$0.35  &  6501.6  &  ---- & ----  &   3.5  \\
EC~12245-2211 &    4.1  &  $-$0.34  &  6140.4  &  $+$0.50 & $-$144.6  &   2.6  \\
EC~12418-3240&    4.1  &  $-$0.34  &  6129.2  &  $+$0.66 & 440.2  &   3.1  \\
EC~12473-1945a&    4.0  &  $-$0.15  &  6246.4  &  ---- &  ---- &    3.0  \\
EC~12473-1945b&    4.0  &  $-$0.10  &  6138.4  &  ---- &  ---- &    2.8  \\
EC~12477-1711 &    4.4  &  $-$0.31  &  6527.2  &  ---- & ----  &   3.5  \\
EC~12477-1724a&    4.2  &  $-$0.24  &  6517.6  &  ---- & ----  &   2.6  \\
EC~12477-1724b&    4.5  &  $-$0.26  &  6497.6  &  ---- & ----  &   3.1   \\
EC~12493-2149 &    4.8  &  $-$0.37  &  6145.6  &  $+$0.65 & 423.6  &   5.2 \\
EC~13042-2740 &   4.6  &  $-$1.85  & 6405.6  &   $+$0.52&   202.6&     5.5*  \\
EC~13390-2246 &    4.8  &  $-$0.36  &  6408.8  &  ---- & ----  &   3.9  \\
EC~13478-2052a&    4.1  &  $-$0.61  & 5420.4  &  ---- & ----  &   5.3   \\
EC~13478-2052b&    4.3  &  $-$0.57  & 5234.4  &  ---- & ----  &   5.2   \\
EC~13499-2204 &   4.4  &  $-$0.65  & 6345.2  &   $+$0.51&   102.2&     5.6*   \\
EC~13501-1758 &    4.2  &  $-$0.15  &  5847.6  &  $+$0.72 &  343.6 &    4.4  \\
EC~13506-1845 &   4.5  &  $-$0.58  & 6664.4  &   $+$0.56&   620.4&     4.9*  \\
EC~13564-2249 &    4.2  &  $-$0.68  & 5903.2  &  $+$0.58 & $-$65.8  &   4.7   \\
EC~13567-2235 &    4.0  &  $-$0.24  &  6341.6  &  $+$0.53 & 179.6  &   2.9   \\
EC~14017-1750 &    4.5  &  $-$1.07  & 6073.6  &   $+$0.63&  283.6 &    5.4  \\
EC~16477-0096  &   3.6  &  $-$2.14  & 4843.6  &   ----&  ---- &    5.6  \\
EC~22874-0038  &   4.0  &  $-$2.41  & 5416.0  &   ----&  ---- &    3.4   \\
BS~16473-0045 &    4.4  &  $-$0.93  & 5356.4  &  ---- & ----  &   5.3   \\
BS~16926-0070  &   4.2  &  $-$1.96  & 5995.6  &   ----&  ---- &    5.3  \\
BS~16469-0074 &    4.5  &  $-$0.44  &  6351.2  &  ---- & ----  &   3.4  \\
BS~16474-0054  &   4.2  &  $-$2.09  & 5570.8  &   ----&  ---- &    4.9  \\
BS~16085-0018 &    3.0  &  $-$1.61  & 5554.0  &  ---- & ----  &   2.2   \\
BS~16085-0004  &   3.7  &  $-$2.11  & 4644.0  &   ----&  ---- &    5.6   \\
BS~16085-0056 &    4.8  &  $-$0.32  &  5220.8  &   ----&  ---- &    5.3  \\
BS~16543-0114 &    3.9  &   0.19  &  4735.2  &   ----&  ---- &    4.7  \\
BS~16479-0031 &    4.3  &  $-$0.22  &  5254.8  &   ----&  ---- &    4.0  \\
BS~16543-0054a&    4.4  &  $-$0.39  &  5747.2  &   ----&  ---- &    4.7  \\
BS~16543-0054b&    4.5  &  $-$0.30  &  5736.8  &   ----&  ---- &    5.1  \\
BS~16477-0078 &    4.6  &  $-$0.11  &  5623.6  &  ---- &  ---- &   5.5  \\
BS~16559-0066 &    4.5  &  $-$0.81  & 4656.4  &  ---- & ----  &   5.7   \\
BS~16551-0015 &    4.8  &  $-$0.55  & 7972.0  &  ---- & ----  &   1.2   \\
BS~16084-0019 &    4.5  &  $-$1.21  & 5998.8  &  ---- & ----  &   4.2   \\
BS~16084-0042 &   4.5  &  $-$0.86  & 7359.2  &   ----&   ----&     5.3*   \\
BS~16087-0004 &   4.7  &  $-$0.63  & 6692.4  &   ----&   ----&     5.4*   \\
CS~22884-0005  &   4.0  &  $-$1.65  & 5558.8  &   $+$0.67&  $-$98.2 &    4.1  \\
G~195-28       &   4.6  &  $-$1.45  & 4698.8  &   $+$0.93&  $-$290.2 &    5.3  \\
G~53-24       &    4.3  &  $-$0.32  &  5281.6  &  $+$0.94& 316.6  &   5.1  \\
G~96-14        &   4.4  &  $-$2.17  & 4562.8  &   $+$1.0&  $-$277.2 &    5.6  \\
G~108-33       &   3.8  &  $-$1.71  & 6226.0  &   ----&  ---- &    0.3  \\
G~115-1       &    4.1  &  $-$0.37  &  5510.0  & $+$0.90 & 457  &   4.8  \\
G~149-34      &    4.9  &   0.32  & 6885.6  &  $+$0.90 & 1832.6  &   0.7   \\
HD~31964      &    1.5  &  $-$0.11  &  6108.8  &  $+$0.55&  43.8 &   $-$0.2  \\
HD~41704      &    4.3  &  $-$0.74  & 5669.2  &  $+$0.50 & $-$615.8  &   5.0   \\
SAO~61681      &    4.4  &   0.28  &  5761.6  &   $+$0.652&  45.6 &    4.8  \\
HD~65934      &    3.0  &  $-$0.04  &  5056.4  &  $+$0.93 &  67.4 &   2.8  \\
HD~89025      &    3.4  &   0.06  & 7255.2  &  $+$0.30& $-$14.8  &  $-$1.1   \\
HD~90861a     &    2.4  &  $-$0.06  &  4732.8  & $+$1.15&  292.8 &    1.4   \\
HD~90861b     &    2.0  &  $-$0.32  &  4572.0  & $+$1.15&  132 &    0.9  \\
HD~90861c     &    2.4  &  $-$0.13  &  4712.0  &  $+$1.15&  272 &    1.4   \\
HD~92588      &    3.5  &   0.08  &  5140.4  &  $+$0.90&  120.4 &    3.8  \\
\hline
\end{longtable}
}
For a few objects more than one spectrum was available as indicated
by symbols a, b, and c, the difference in estimated values is indicative
 of the internal error.
$*$ The M$_V$ for hot metal-poor stars is uncertain because we did not have
  good calibrators covering that temperature and  metallicity range.
}}

\section{ Calibrated stars}

\noindent
We have initiated a program for the definite identification of  metal-poor
candidates  from different surveys such as  the objective prism  survey of \cite{be92}, which is generally referred to as the HK survey, 
the Edinburgh $-$ Cape blue-object survey by \cite{st97}, and the high
 tangential velocity objects listed by \cite{le84}. During 1999-2001 we
  obtained spectra of a modest sample and also a good number of
 stars of known parameters.  The semi-empirical
approach adopted in \cite{gi02} resulted in identifying 
 and parametrizing the metal-poor star candidates at a very slow pace, hence
 we chose to explore an ANN-based approach. 
 Our earlier attempt at using the spectra of calibrated stars from the 
known empirical library (e.g. \cite{ja84} for training the
network and then employing them for parametrizing our sample proved to be difficult despite  
our attempts at matching the resolution of 
two spectra. We faced convergence problems, and the calibration errors 
were unacceptably large. The spectral libraries available then
 also had no stars with good coverage in metallicity.

 On the other hand, using stellar spectra of calibrated stars obtained with
 the same instrument configuration and comprising stars evenly distributed in
 parameter space yielded a very good calibration accuracy even for calibrated 
 samples of modest size. It should be noted that the spectral resolution
 and spectral coverage of our spectra are well suited for our objective.

We therefore created a library of observed stellar spectra for stars 
with well-determined parameters (adding more spectra in
 2004-06), which was used for training ANNs. These were
 used to estimate the astrophysical parameters, 
$\rm T_{eff}$, log~$g$, \feh, and $M_V$ for a modest sample
of unexplored field stars  using medium-resolution stellar spectra.

Our database of stars with known spectral classification and parallaxes is
presented in Table 1, which contains the star name, the Hipparcos number, the V magnitude, 
(B$-$V), log~$g$, $T_{\rm eff}$, \feh, and references for the stellar
\feh. Many objects were observed more than once. These objects with known 
atmospheric parameters were selected primarily from \cite{gr01}, \cite{al99}, \cite{sn01}, and
\cite{ca01}. \cite{gr01} have calculated 
atmospheric parameters with the following uncertainties: 80 K in T$_{eff}$, 
0.1 in log $g$, and 0.1 in [M/H]. The temperatures tabulated by
 \cite{al99} have an uncertainty of \mbox{200 K},
 while the uncertainty in log~$g$ varies from $\pm0.1$ at log $g$ of 4.5 to 
as much as $\pm$0.5 at log~$g$ of 2.2. The uncertainties in the \cite{sn01} data are the following: 150 K in T$_{eff}$, 0.3 in log~$g$, and
0.2 in \feh.  We also made use of the \feh derived from the high-resolution
spectroscopy of the individual stars available in literature and those 
from the Elodie data base (\cite{so98}, whose  
parameter uncertainties are 145 K in T$_{eff}$, 0.3 in log~$g$, and  
0.2 in \feh. The 73 metallicity calibration  stars with known log~$g$, 
$T_{\rm eff}$,  and \feh contained in Table 1 (full table available in electronic form) 
are indicated  with an asterisk mark.
We observed more than two hundred stars and rejected those with binarity
or other peculiarities such as Ap-Am spectra and those with
emission lines. Some 
spectra were rejected due to poor signal to noise (S/N) ratios.

\section{Observation and data handling}
The spectra were obtained using a medium-resolution Cassegrain spectrograph 
mounted on the 2.3 m Vainu Bappu Telescope at VBO, Kavalur, India. When 
used with a grating of 600 grooves mm$^{-1}$ and a camera of  1500 mm focal 
length, the spectrograph gives an average dispersion of 2.6 \AA\ per pixel.
During the extended period of several years, over 200 medium-resolution spectra  
were obtained. The spectral coverage is 3800--6000 \AA. The spectra were 
recorded on a 1K $\times$ 1K CCD (with Thomson TH77883) with a pixel size of
24 $\mu$. The setup gave a two-pixel resolution of 1200.

The reduction and analysis of the spectroscopic data were performed using the
standard spectroscopic packages in IRAF.
  All CCD frames were bias-corrected, response-calibrated
using dome-flat spectra, and cleaned for cosmic rays.  
Even before converting them to wavelength scale, the extracted spectra 
 were aligned accurately using a script to ensure that a given
 spectral feature fell on the same pixel number in all spectra. This 
 procedure has the disadvantage that radial velocity information is
 not retrieved. No absolute flux calibration was performed.
 For fainter stars 2$-$3 exposures were combined to attain an S/N ratio
 of atleast 50.
 For the continuum-fitting we adopted a procedure similar to that
 given in \cite{sn01}.
 The spectra exhibiting emission lines were excluded from the sample.
 The  spectra were trimmed such that all  spectra (700 pixels)
 covered exactly the same spectral region. The alignment of the spectra
 is crucial to obtain the desired accuracy. 
  Fig. 1  shows representative stars from our sample; the stars  
with near-solar metallicity arranged in decreasing temperature sequence
from top to bottom. We superposed the spectra of metal-poor stars
with similar temperature  and show their atmospheric parameters \te, log~$g$, and \feh within parenthesis.

\section{Atmospheric parameters of the training set }
Table~1 lists the atmospheric parameters \feh,
$T_{eff}$, and log $g$ compiled from the literature and adopted for each star
in our study.  We took
particular care to select stars that span a wide range in
\feh, $T_{eff}$, and log~$g$;
these values were used to train the  ANNs; 
the reference for  \feh is given in the last column of Table 1.
These [Fe/H] are estimated using high-resolution spectra and
 model atmospheres, hence their accuracy probably is about
$\pm$0.2 dex. The stars used for metallicity correction
 (with known \feh, $T_{eff}$, and log~$g$)
 are indicated by an asterisk following the star name.
We used the back-propagation ANN code  developed by B.D.Ripley 
(see \cite{ri93}, \cite{ri94}.  The ANN configuration is same as that employed 
in our earlier work (\cite{gi06}). 
Separate ANNs were trained for each parameter.

\section{ANN atmospheric parameter results}

\subsection{Metallicity, [Fe/H]}
The top panel in Fig.~ 2 shows a plot of the \feh residuals obtained using the  ANN for 
the 73 stars against their \feh  taken from the literature and given in 
Table~1. The [Fe/H] metallicities range from $-$3.0 dex to 0.3 dex, and
the reduced mean scatter about the line of unity is 0.15 dex. The [Fe/H] 
estimates quoted in the literature often have uncertainties in the range 
0.2$-$0.4 dex. To test the goodness of the ANN, we divided the sample into 
two parts and trained the network separately on each part. Then the weights 
of ANN trained for  part 1 were used to estimate \feh for the stars in 
part 2. The middle and bottom panels of Fig. 2 show an rms error of 0.31 and 0.22,
which is indicative of the accuracy with which the ANN can predict the 
metallicity of a given star within the trained metallicity range.

Using the weights from the ANN trained for the sample of calibrated stars, 
the metallicity of the candidate metal-poor stars could  be estimated. These estimates
were subjected to independent tests to avoid higher temperature -- low-metallicity 
degeneracy, but they were still useful
 in segregating the stars of near-solar
metallicity (\feh in $-$0.5 to $+$0.3 dex range)~ from significantly metal-poor 
objects with \feh $<$ $-$0.5 dex.

\begin{figure*}
\centering
\includegraphics[angle=0,height=12cm,width=12cm]{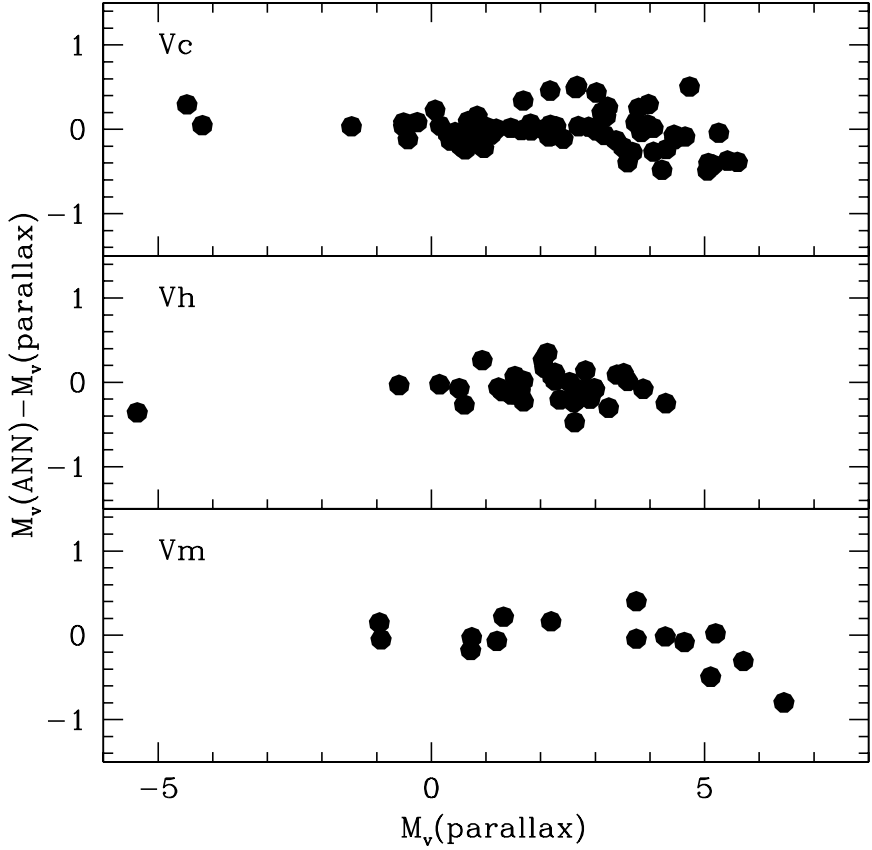}
\label{bop}%
\caption{Plot of M$_{V}$$_{ANN}$ $-$ M$_{V}$$_{Par}$ 
 versus M$_{v}$ (parallax). The subset of cool stars (Vc) is
plotted in the top; the middle panel contains the results for hot stars (Vh), 
and the bottom panel the results for metal-poor stars (Vm). }
\label{Figure 4}
\end{figure*}

\subsection{Temperature and surface gravity}

The literature contains a larger  number of stars with good estimates of
 \te\ and  log~$g$  values compared with those with  \feh values.
For stars with near-solar metallicity we used temperatures and gravities 
 given in \cite{al99} and \cite{gr01} for temperature and calibration.
 For metal-poor stars \te~ and log~$g$ were mostly taken from
 \cite{sn01}. In Fig. 3 we plotted these parameters for the common
 stars to estimate systematic differences between the two works.
We found that the T$_{eff}$ values obtained by  \cite{gr01} are
systematically higher by about 1.5\% and for log~$g$ the systematic difference
 is 3\% to 4\%.
Hence we believe that our compiled calibrating set is not affected by large systematic errors.

On the other hand,
it should be noted that the  metal-poor stars do not have strong features 
in their spectra because of their lack of metals, while hotter stars lack 
strong metallic features in their spectra because of ionization. To 
ensure that the ANN does not become confused by this, we divided the stars into
solar metallicity (\feh $>$ $-$0.5 dex) and metal-poor (\feh $<$ $-$0.5 dex) 
groups.

 We estimated the
\feh for the sample stars in  Table 1 with known \te~ and  log~$g$ using the ANN trained 
for \feh as explained in section 5.1. We have spectra of 110 
calibrated  stars with near-solar metallicity and spectra of 33 metal-poor 
stars. We trained the temperature ANN separately for each metallicity group.

The solar-metallicity stars were separated into two random groups of 55, 
and a sanity check similar to that demonstrated in Fig. 2 was 
performed.  The rms about the line of unity was found to be 150 K for
both groups.   As the temperatures found in the literature have 
errors that can be as high as 200 K, this is not surprising.

  We trained two ANNs for log~$g$, one ANN with stars with 
\feh $<$ $-$0.5 dex, and the other with \feh $>$ $-$0.5 dex. 
A procedure similar to that given for \te\ was adopted.
  The accuracy of the log~$g$ estimate is in the range 0.3 to 0.5.

\begin{figure*}
\centering
\includegraphics[angle=0,height=12cm,width=12cm]{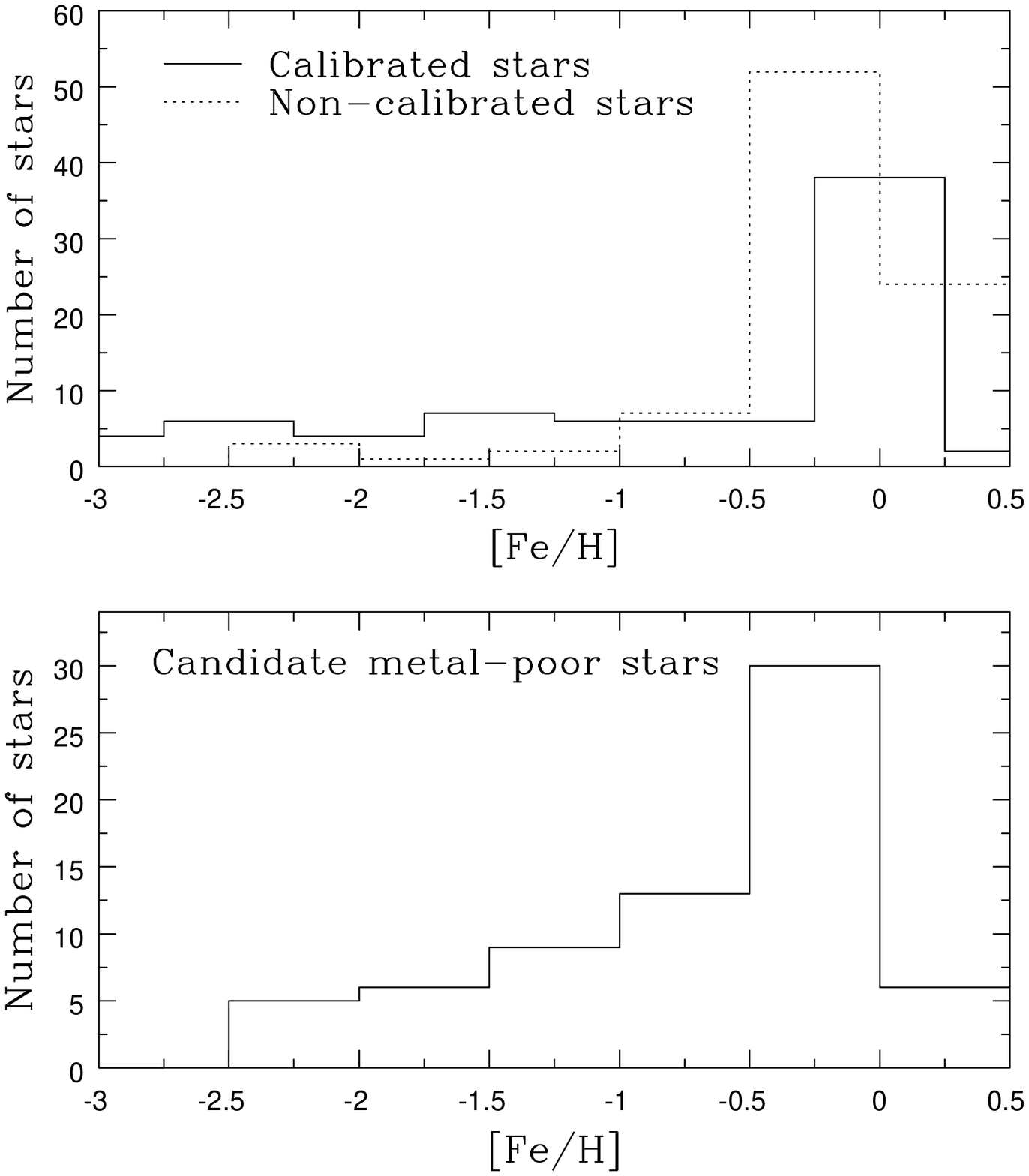}
\caption{Metallicity distribution for the calibrated sample and
 candidate metal-poor stars.
}
\label{Figure 5}
\end{figure*}

\onlfig{6}{
\begin{figure*}
\centering
\includegraphics[bb= 50 326 592 635]{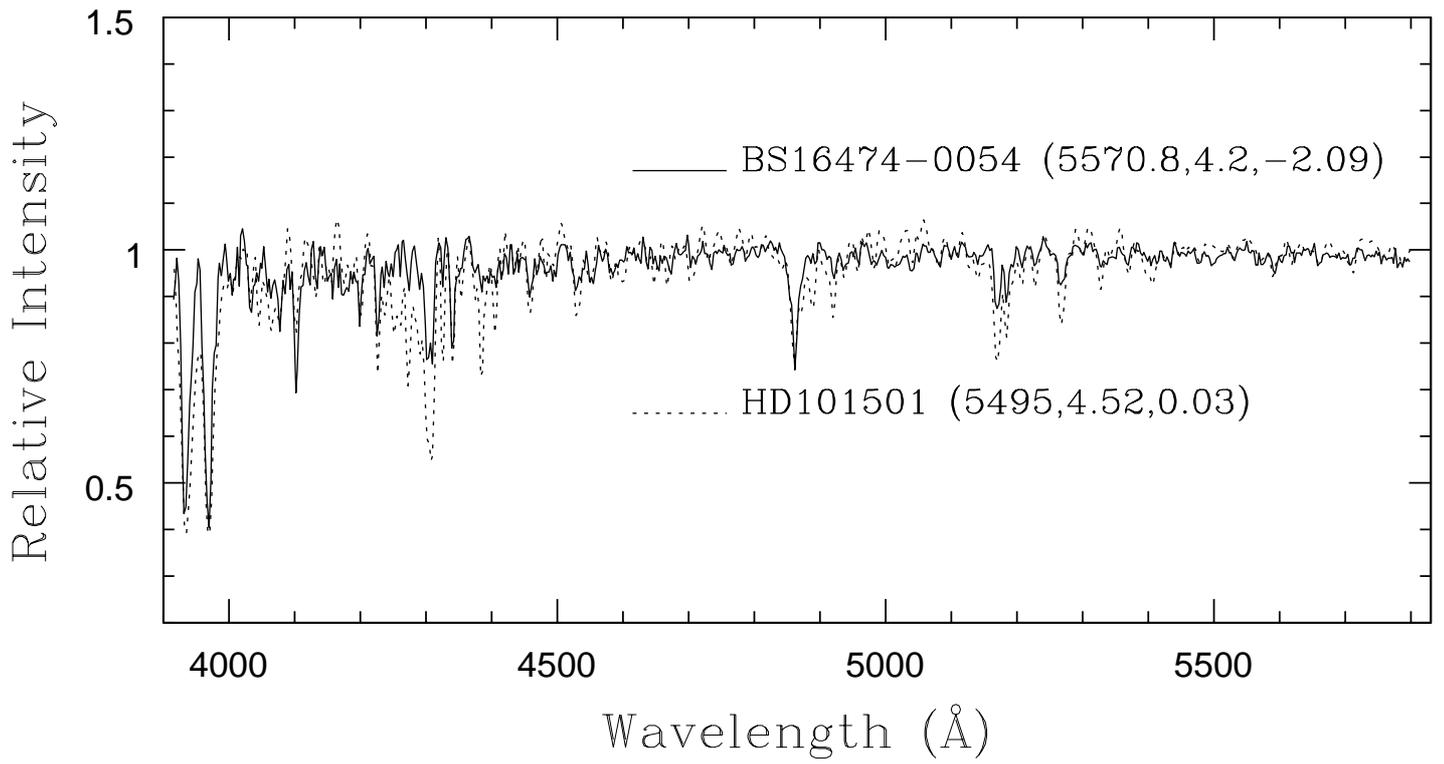}
\caption{Spectra of metal-poor stars compared with stars of similar temperature, 
gravity, and near-solar composition. The solid line indicates the metal-poor 
stars and the dashed line indicates solar-metallicity stars. The atmospheric parameters \te, log $g$, and \feh for
 each star are given in parenthesis.}
\end{figure*}}

\begin{figure*}
\centering
\includegraphics[angle=0,height=12cm,width=12cm]{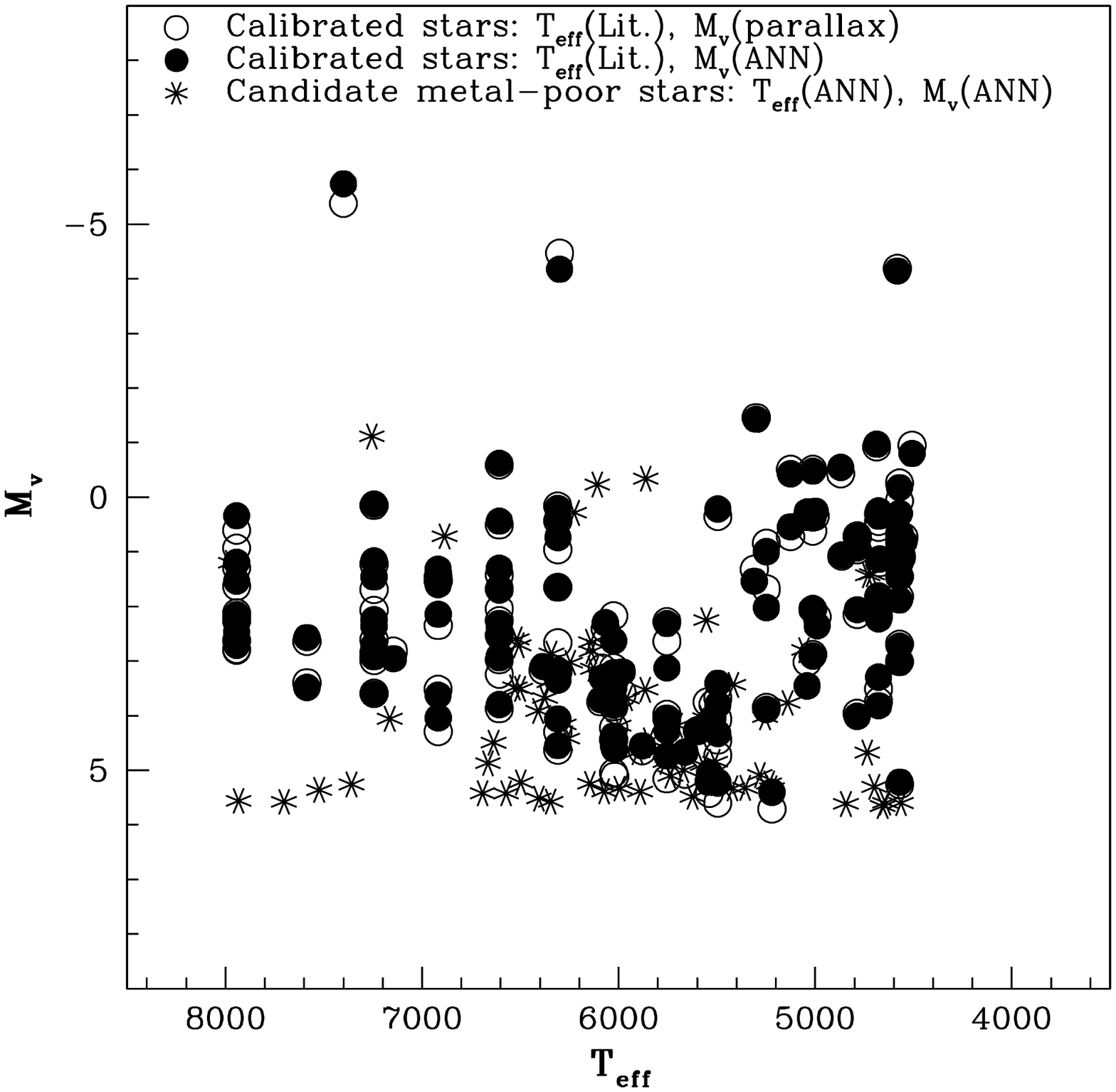}
\caption{ M$_{v}$ as a function of T$_{eff}$ for
 the luminosity-calibration stars and candidate metal-poor stars.
}
\label{Figure 5}
\end{figure*}

\section{ANN absolute magnitude results}
 A large  portion of the 
stars observed by us have parallax estimates.  Combining the 
$V$-magnitudes with the Hipparcos parallaxes and absolute $V$-magnitudes, the 
$M_V$ could be calculated.  Stars with a parallax error greater than 20\%
were excluded.  Since most of the sample stars are nearby bright stars, 
the effect of interstellar reddening  in most cases will be weak or even
negligible. We therefore excluded this correction from our $M_V$ 
calculations. Our spectral region contains many luminosity-sensitive 
features such as wings of hydrogen lines, lines of Fe {\sc ii},
Ti {\sc ii}, Mg {\sc i} lines at 5172$-$83 \AA,  and for later spectral types
G-band blends of MgH, TiO, VO, etc.
 However, the same features cannot serve the  whole range of 
spectral types; hence, we divided the sample stars into two groups based 
upon their temperatures. Group I had stars in the temperature range 4300$-$6300 K
labeled Vc and 
group II those in the 6600$-$8000 K range labeled Vh. Yet another group, group III, which contains 
metal-poor stars Vm, was handled separately. The stars in this group 
have temperatures in a range similar to that of group I.

 Fig. 4 illustrates the errors  associated with these three groups. 
The ANN trained for group I (with 76 stars labeled Vc in figure) could predict $M_V$ with an
accuracy of 0.22 mag, while the ANN for group II (with 39 stars) attained an 
accuracy of 0.18. The group III of metal-poor stars had a very few stars (14) 
and could predict $M_V$ with an accuracy of only 0.29.
  An error of 0.3 mag in luminosity would result in an error
 of 150 parsec in distance at a distance of 1 kpc. 
 One likely reason for  $M_V$  error could be that luminosity sensitive features
 like lines of the Fe {\sc ii},
Ti {\sc ii}, and Mg {\sc i} lines at 5172$-$83 \AA~ are also  
 metallicity dependent. Furthermore, the number of metal-poor
 stars with good parallaxes is woefully small. The large systematic error for low-luminosity objects with M$_V$ of 5 deserves to
be analysed with additional data. Another possible solution is the usage of line ratios appropriate
 for metal-poor stars, as suggested by \cite{co87} and \cite{gr89}.

\section{ Stellar parameters for the candidate  metal-poor stars}

The metallicity distribution of the
 calibrated sample is presented in the top panel of Fig. 5.
 The figure shows the distribution of the stars with [Fe/H] taken from the 
 literature with a thick continuous line. An additional  110 stars with  well-determined
 \te\  and log$~$ g were lacking good [Fe/H] estimates. We 
  determined [Fe/H] for these objects using the ANN, and their metallicity
 distribution is  presented with  a dotted line. 
 The distribution shows that we had a good coverage of training
 stars in different metallicity bins.
 
We observed candidate metal-poor stars from the
objective prism  surveys of \cite{be92} (BPS),
the Edinburgh $-$ Cape blue-object survey (EC) by \cite{st97}, and the high
 tangential velocity objects listed by \cite{le84}.
 We also included some unexplored high proper motion field stars.
 Using three separate ANNs, we estimated atmospheric parameters 
for the candidate metal-poor  stars. At first, the metallicity was estimated 
using an ANN trained for metallicity. This helped us in separating the metal-poor 
stars from those of near-solar metallicity or moderately metal-poor objects.
A separate ANN trained for these two groups was employed to estimate the 
\te\ and log~$g$ for these stars.
 The estimated atmospheric parameters are presented in Table 2. A few stars 
had more than one spectrum and the small differences between the
 parameters estimated from each spectrum are indicative of internal errors. 
 The (B$-$V) colors were available in SIMBAD for 
many of them, which were used to verify the temperatures estimated by the ANN.
 We used the calibration tables of \cite{sc82} to estimate the 
 photometric temperatures.
We tabulated the difference between \te (ANN) and \te (photometric)  in Table 2.
 We obtained surprisingly high residuals for EC 11175-3214, EC 11260-2413, EC 13506-1845, 
 and G 149-34. While the observed spectrum strongly supports the \te~estimated
 from the ANN, a misidentification  cannot be ruled out.
 Excluding these exceptions, residuals indicate an rms error of 265 K.
  Many metal-poor candidate stars were near the faint limit, hence the S/N ratio was in the range
 of 40-50, while most of the calibrated star spectra had an S/N ratio higher than 100.

 Within our modest sample of  stars, a good fraction (about 20\%)
 are significantly metal-poor with [Fe/H] in $-$1.0 to $-$2.5 range.
 We find that 33\% of the BPS stars and 21\% of the EC stars 
 belong to the [Fe/H] range of  $-$1.0 to $-$2.5. A few high
   proper motion Giclas objects studied also contain metal-poor stars, but 
  the number studied is currently  very small, therefore we do not offer statistics. 

 The bottom panel of Fig. 5 shows the metallicity distribution of candidate metal-poor stars,
 which shows that our candidate sample has a large portion of moderately
 metal-poor stars, but the fraction of significantly metal-poor 
 star is also encouraging.

We have plotted in Fig. 6,  a newly identified metal-poor star, BS 16474-0054 along with a near-solar-metalicity 
star of similar temperature to substantiate our findings.

 With these encouraging results (notwithstanding the small sample)
 we propose to extend this work to a much larger sample of candidate 
 metal-poor stars from surveys such as the HK II decribed in \cite{be05}.

  With the help of the estimated \te\ and M$_V$, 
we are able to place the program stars in the H-R diagram, as shown in Fig. 7.  
 The luminosity-calibration stars with M$_V$ taken from
 the literature are shown by open circles; their 
  M$_{V}$ estimated from the ANN is shown by a filled circle.
 The difference between the two is indicative of internal errors.
 In both cases the \te\ is the catalog value.
 It should be noted that our calibrated stars do not represent the 
local neighborhood alone since the objects were taken from different sources
 to encompass the required range of stellar parameters (metallicity in
 particular). Hence the \te\  and M$_{V}$ diagram has a large scatter even for
 the calibrated stars.

 A good fraction of candidate metal-poor stars are dwarfs or subgiants
 (which possibly are slowly evolving low-mass stars) although the calibrated
 stars in Table 1 also contain several giants  among the significantly
 metal-poor stars.

\subsection {Limitation of our approach and future strategy}

  We are aware of the problems caused by degeneracies in certain
 parameter domains. We avoided these sources of inaccuracies by 
  incorporating a branching procedure that resulted in the segregation of
 data into meaningful subgroups. This additional step considerably
 improved the accuracies of the derived parameters compared with our
 earlier work (\cite{gi06}).

 It should be noted that our ANN-based approach does not 
 allow for extrapolation. For example, the [Fe/H] network is 
 trained for a [Fe/H] range of $-$3.0 to $+$0.3 and therefore may not 
 give reliable results for super-metal-rich stars or Ap-Am stars.
 This approach is also not applicable for double-line spectroscopic
 binaries.

 As mentioned earlier, the ANN procedure adopted here is not suitable
 for handling peculiar stars; however, it does provide a good estimate of 
 \te, log~$g$, and metallicty for candidate metal-poor stars from the
 surveys mentioned previously. The \te\ estimated here  agree  with
 those estimated from (B$-$V)  within $\pm$265 K for
 candidate stars with the exception of a few  stars.

 Although this maiden effort of estimating M$_V$ from spectral features
 is quite accurate for solar metallicity objects, the errors are
 large particularly in the low-luminosity regime for the metal-poor stars. 
In addition to full spectra, we propose to input 
 some important line ratios and explore near-IR features in our future work.

\section{ Summary and conclusions}

 We have developed an empirical library of stellar spectra for stars
  covering a temperature range of 4200 $<$ \te  $<$ 8000 K,
  a gravity range 0.5 $<$ log~$g$ $<$ 5.0, 
  and a metallicity range of $-$3.0 $<$ \feh $<$ +0.3. With the  good 
  spectral coverage of 3800--6000 \AA, several spectral features showing
  strong sensitivity to the stellar parameters were available, which
  were used by the ANN in the learning process. 

  The procedure of pre-classifying the data and training separate ANN for
  each subgroup resulted in a significant increase in accuracies.
  Now temperatures could be estimated within $\pm$150 K.
  Similarly, using of three separate ANNs for hot, cool, and
  metal-poor stars yielded a very good accuracy in  M$_{V}$ 
  calibration. 

    We used these trained networks primarily to detect metal-poor
   objects  from  a modest sample of unexplored objects. However,
   the empirical library developed may be useful for other 
   applications  and can  be accessed by interested users on request. 
   We believe that it will be useful in studying stellar population
   in large samples of galactic stars.

    We extended the application of ANN to M$_{V}$ with an
   accuracy of $\pm$0.3 dex. The primary application of 
   M$_{V}$ is in distance determination, and the spectroscopic approach
   based upon the strength and profiles of the lines is independent of 
   reddening. In addition, the M$_{V}$ calibration 
   can be used for the quick
   identification of objects of various luminosity types in
   large databases containing heterogeneous objects.

   {\bf Future prospects}
\noindent 

   With the ANN procedure giving the desired accuracy established here, 
we would like to explore a much larger sample of 
   candidate metal-poor stars. We also need to 
    extend  the empirical library toward hotter 
   temperatures and also overcome the poor coverage of low-gravity
   objects. We also contemplate including near-IR
   O {\sc i} feature, Ca {\sc ii} lines, and  line ratios suggested
   by \cite{co87} and \cite{gr89} for the luminosity 
   calibration of metal-poor stars.
   In this preliminary work, we used (for calibration)
   M$_V$ for  nearby stars estimated from the parallaxes omitting 
   reddening corrections. Better and enlarged samples of M$_{V}$ from
   upcoming surveys or data releases of the ongoing survey could
   be used to attain consistent M$_{V}$ accuracy in the full temperature
   range, which will help in understanding the evolutionary
   status of the candidate stars.

    We have an ambitious project of observing an extended 
   sample of F-G stars covering  a broad range in galactocentric
   distance to study the metallicity gradient, which is known to
   exhibit two slopes and also some wriggles near the spiral arm 
   locations. We hope to attain the required accuracy in metallicity
    by carefully binning the data in a more narrow range in
   temperatures and gravities, and also including important line
   ratios.

{\it Acknowledgments}\\
\noindent
Sunetra Giridhar wishes to thank T. Van Hippel for his help with the ANN 
code. 
This work was partially funded by the National Science Foundation’s Office of International
Science and Education, Grant Number 0554111: International Research Experience for Students,
and managed by the National Solar Observatory’s Global Oscillation Network Group.
This work 
made use of the SIMBAD astronomical database, operated at CDS, Strasbourg, 
France, and the NASA ADS, USA. We thank the anonymous referees for their constructive comments, which helped us
to improve the manuscript.

\end{document}